# Stability Enhancement of Cascaded DC Systems via Source Output-Impedance Reshaping Using Model-Free Control

Wanrong Li
*School of Electrical and Computer Engineering*
*The University of Sydney*
Sydney, Australia
wanrong.li@sydney.edu.au

Shuhan Hu
*School of Electrical and Computer Engineering*
*The University of Sydney*
Sydney, Australia
shuhan.hu@sydney.edu.au

Hengping Xiong
*School of Electrical and Computer Engineering*
*The University of Sydney*
Sydney, Australia
hxio0294@uni.sydney.edu.au

Runnan Wang
*School of Electrical and Computer Engineering*
*The University of Sydney*
Sydney, Australia
runnan.wang@sydney.edu.au

Huawei Yuan
*Department of Electrical Engineering*
*City University of Hong Kong*
Hong Kong, China
huawyuan@cityu.edu

Jianguo Zhu
*School of Electrical and Computer Engineering*
*The University of Sydney*
Sydney, Australia
jianguo.zhu@sydney.edu.au

Sinan Li
*School of Electrical and Computer Engineering*
*The University of Sydney*
Sydney, Australia
sinan.li@sydney.edu.au

***Abstract*—Cascaded DC systems may become unstable through source–load impedance interactions even when their constituent converters are locally stable. This paper proposes a source-side model-free control (MFC) method that reuses bus-voltage feedback to reshape the source output impedance while preserving nominal reference-to-output dynamics. The controller uses only nominal source-loop information and requires neither a load-converter model nor load-side measurements. A current-controlled Buck source supplying a regulated Buck–resistor load that exhibits constant-power load (CPL) behavior at low frequencies is used as the case study. Impedance analysis predicts that the proposed MFC increases the maximum stable load power by approximately 50%, thereby improving the stability margin of the cascaded system. Switched-model simulations confirm preservation of the nominal reference-to-output response and show that the MFC-controlled system remains stable at load levels for which the baseline system becomes unstable.**

## I. Introduction

Modular dc distribution systems commonly cascade a bus-regulating source converter with a tightly regulated downstream converter, as illustrated in Fig. 1 [1]. Although each converter may be stable independently, the regulated load can exhibit negative incremental input impedance at low frequencies, allowing source-load interaction to destabilize the dc bus [2]. Following Middlebrook's impedance criterion, $|Z_s(j\omega)|<|Z_L(j\omega)|$ at all frequencies is a conservative sufficient stability condition when the isolated converters are stable [3], [4]. Here $Z_s$ and $Z_L$ are source output and load input impedances; their ratio $L_m= Z_s/Z_L$ is the impedance minor-loop gain.

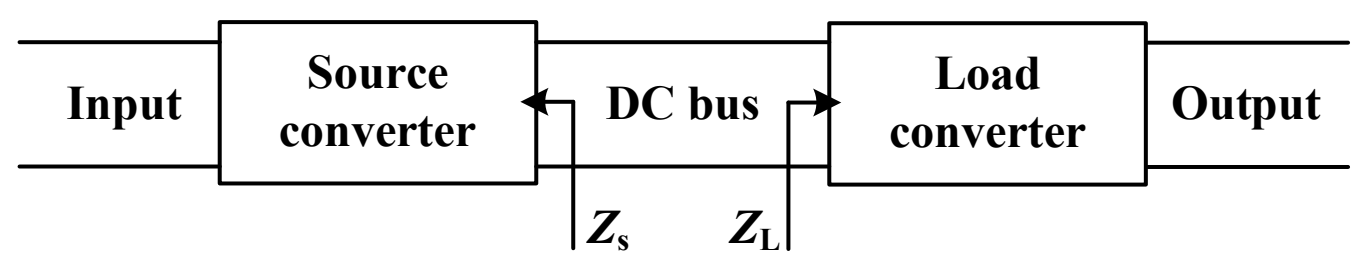

Fig. 1. Basic architecture of a cascaded dc distribution system.

Stabilization therefore seeks adequate source–load impedance separation. Resistor-based passive damping requires additional components and may incur power dissipation. Negative-input-resistance compensation and virtual impedance reshape the load port but require access to the load controller [5], [6]. Existing source-side active-damping methods introduce a damping or impedance-shaping block whose parameters are selected for a desired damping or impedance profile [7]–[9]; a power-adaptive variant additionally use operating-power information [10]. However, these approaches either require load-side information or access to the load controller, or do not explicitly guarantee preservation of the baseline source reference-to-output dynamics. Preserving these dynamics allows the source output impedance to be reshaped for stability without changing the pre-tuned voltage-command response.

Building on the ultra-local-model-based MFC decoupling framework in [11], this paper adds a local bus-voltage feedback loop to the existing source controller. It uses nominal source-loop information but neither a load-converter model nor load-side measurements. The added loop is designed to preserve the nominal reference-to-output transfer function of the source converter, so the source-port impedance can change without nominally changing the reference-to-output dynamics. The proposed method is evaluated in a cascaded Buck system with a voltage-regulated Buck converter supplying a resistive load, whose input port exhibits constant power load (CPL) behavior at low frequencies. Stability boundaries predicted from the converter's closed-loop input impedance are compared with time-domain switching-model simulations. The increase in critical downstream load power, evaluated using identical source power-stage and baseline-control parameters, provides a quantitative measure of the improvement in source–load interconnection stability.

## II. Source-Side Impedance Reshaping

To illustrate the source-side impedance-reshaping principle, Fig. 2(a) gives a small-signal Thevenin-equivalent source-load representation of the cascaded system at a fixed operating point. Figs. 2(b) and 2(c) show the baseline and MFC source-side control configurations, respectively. Here, $T(s)$ is the nominal voltage-loop gain, and $Z_o(s)$ is the open-loop output impedance of the source. $G_{v,base}(s)$ and $Z_{s,base}(s)$ respectively denote the baseline controller-based closed-loop reference-to-output transfer function and the closed-loop source output impedance, which are given by

$$G_{v,\text{base}} = \frac{T}{1+T}, \tag{1}$$

$$Z_{s,\text{base}} = \frac{Z_o}{1+T}, \tag{2}$$

For the MFC source,

$$G_{v,\text{MFC}} = \frac{C_o T}{1+(C_o+1)T}, \tag{3}$$

$$Z_{s,\text{MFC}} = \frac{Z_o}{1+(C_o+1)T}. \tag{4}$$

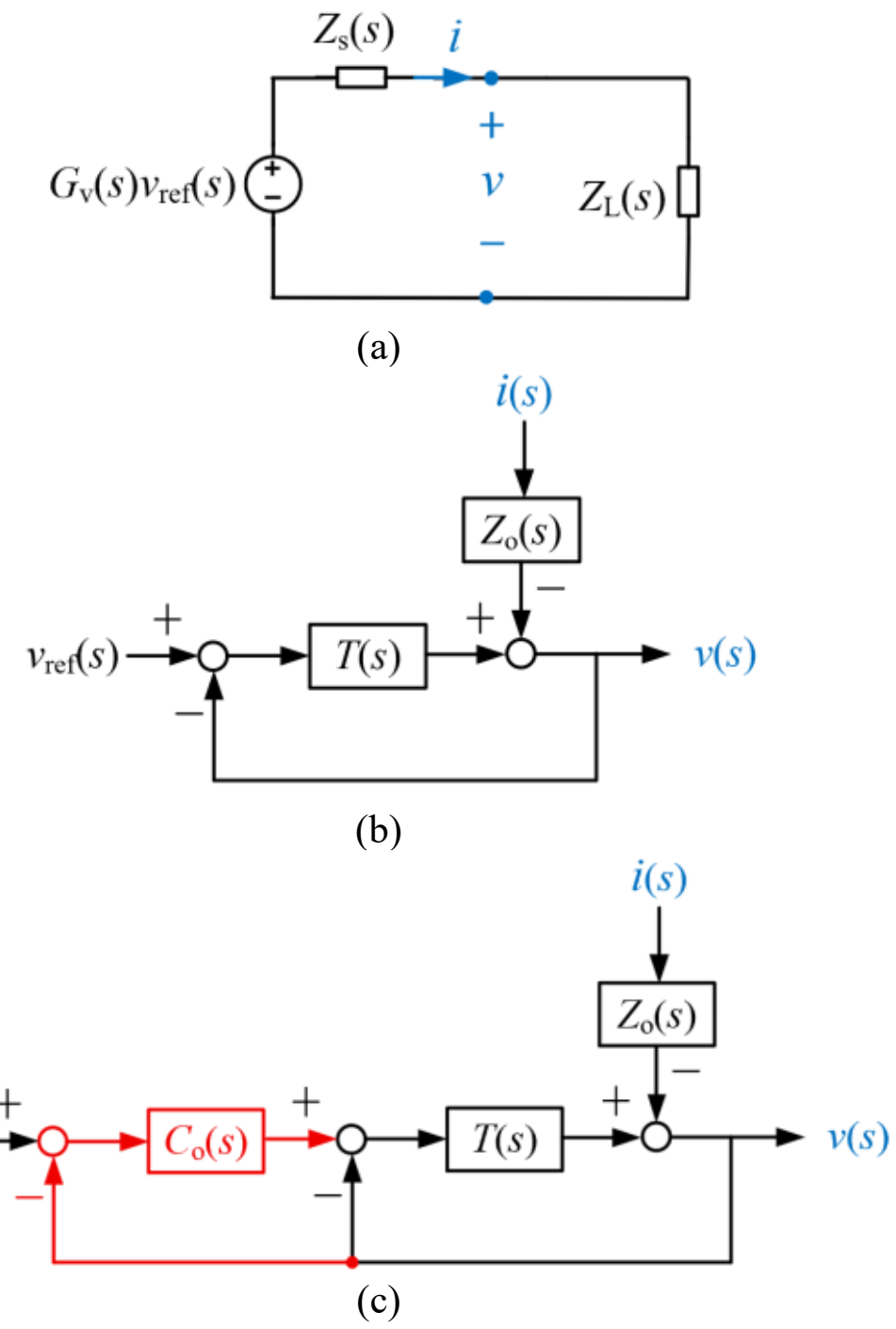

Fig. 2. (a) Source–load Thevenin model; (b) baseline source converter control loop; (c) proposed MFC loop.

To preserve the baseline nominal reference-to-output transfer under zero load-current perturbation, the additional-loop controller $C_o(s)$ in MFC is selected as $C_o(s)=1+T(s)$ so that

$$G_{v,\text{MFC}} = G_{v,\text{base}}, \quad Z_{s,\text{MFC}} = \frac{Z_{s,\text{base}}}{1+T}. \tag{5}$$

From (5), the additional MFC path scales the baseline source output impedance by $1/[1+T(s)]$. Since the load impedance remains unchanged, the source–load minor-loop gain is scaled by the same factor. Consequently, the source output impedance and the minor-loop magnitude are reduced over the frequency range in which $|1+T(j\omega)|>1$. At frequencies well above the source voltage-loop bandwidth, $|T(j\omega)|$ approaches zero, and the MFC-controlled output impedance converges to the baseline output impedance. The additional MFC path uses only nominal source-loop information and the dc-bus voltage already measured by the baseline controller; hence, it requires neither a downstream-converter model nor load-side measurements, online load-power estimation, or additional voltage sensing.

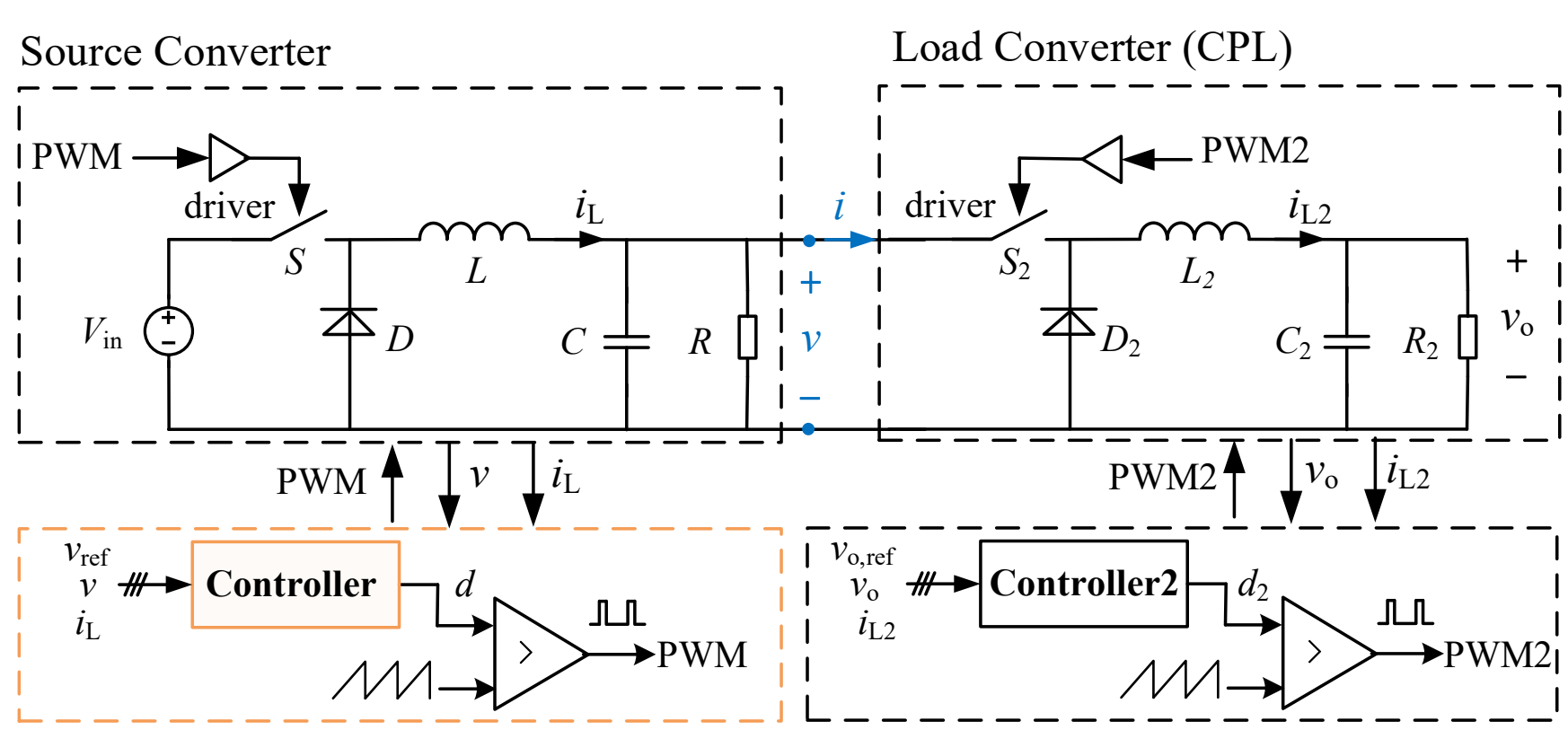

Fig. 3. Circuit and control configuration of the cascaded dc system comprising a Buck source converter and a regulated Buck load converter feeding a resistive load.

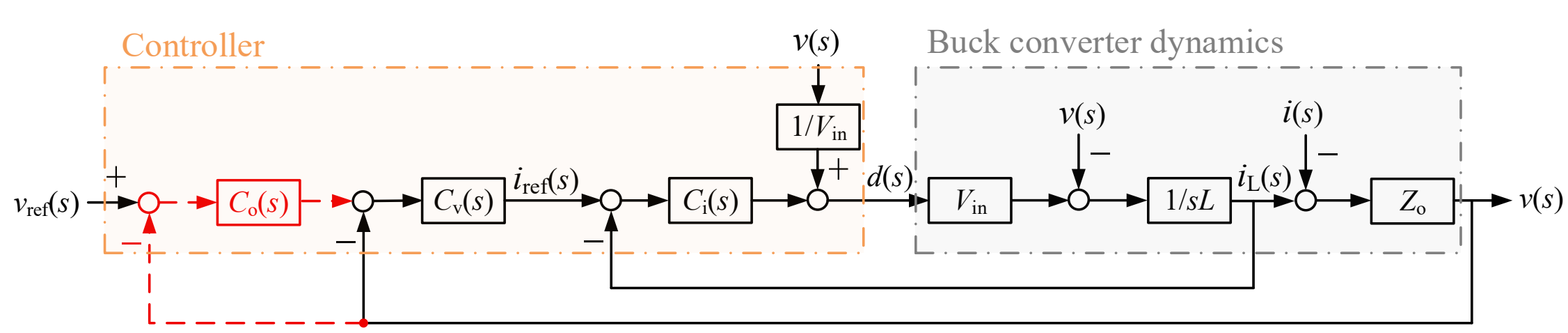

Fig. 4. Averaged source-Buck stage, cascaded voltage–current control, and added MFC additional path.

## III. Cascaded Buck System and Impedance Modeling

Fig. 3 shows the 48-to-24-V Buck source converter, the source-side RC shunt, and a downstream regulated Buck converter feeding a resistive load [2]. The CPL is realized by a voltage-regulated Buck converter feeding a resistive load. Fig. 4 presents the averaged Buck-source model and the two control architectures being compared: the baseline employs cascaded outer-voltage and inner-current loops [12], [13], while the proposed MFC additional path is added for source output-impedance reshaping. To enable a fair comparison, the source power stage and the baseline inner-current and outer-voltage controllers are kept identical in the baseline and MFC cases. The circuit and controller parameters are listed in Table I.

### A. Closed-Loop Output Impedance of the Source Converter

Under continuous-conduction-mode operation, the averaged large-signal dynamics of the source-side Buck converter are

$$L\frac{di_L}{dt} = dV_{\text{in}} - v\,, \tag{6}$$

$$C\frac{dv}{dt} = i_L - \frac{v}{R} - i\,. \tag{7}$$

Here, $L$ is the source-converter output inductance, $C$ is the dc-bus capacitance, $R$ is the resistance of the shunt branch connected across the dc bus, $V_{\text{in}}$ is the source input voltage, $d$ is the averaged duty ratio, $i_L$ is the source-inductor current, $v$ is the dc-bus voltage, and $i$ is the input current drawn by the downstream converter as shown in Fig. 3. For the downstream Buck converter, $d_2$, $i_{L2}$, and $v_o$ denote the duty ratio, inductor current, and output voltage, respectively, and $L_2$, $C_2$, and $R_2$ denote the corresponding inductance, capacitance, and resistive load.

With $v/V_{\text{in}}$ feedforward canceling the bus-voltage term, the inductor's integrating dynamics allow the proportional controller $C_i(s) = K_{pc}$ to yield the nominal first-order closed-loop response $G_i(s) = 1/(1+s\tau_i)$ from $i_{\text{ref}}$ to $i_L$, where $\tau_i = L/(V_{\text{in}}K_{pc})$; an inner-loop integrator is therefore unnecessary for the nominal first-order design. The outer-loop plant is $G_i(s)Z_o(s)$, where $Z_o(s) = R/(1+sRC)$ is the open-loop source output impedance set by the bus RC shunt branch. The outer voltage loop uses conventional PI compensation, $C_v(s) = K_{pv} + K_{iv}/s$ [12]. Placing its zero at $K_{iv}/K_{pv} = 1/(RC)$ cancels the RC-network pole in $Z_o(s)$, leaving an integrator and the inner-loop pole in the loop gain

$$T = C_v G_i Z_o = \frac{\omega_c}{s(1+s\tau_i)}\,, \tag{8}$$

where $\omega_c = K_{iv}R = K_{pv}/C$ sets the nominal outer-loop crossover frequency. The MFC is formed by adding only the $C_o(s)$ block to the baseline controller. The closed-loop source output impedances under the baseline and MFC controllers are given in (2) and (4).

### B. Closed-Loop Input Impedance of the Load Converter

For the regulated Buck load in Fig. 3, the bus and load-output voltages at the nominal operating point are $V_{\text{ref}}$ and $V_{\text{o,ref}}$, respectively. The bus-side input current satisfies $i = d_2 i_{L2}$. Linearizing this relation gives

$$\hat{i} = D_2\hat{i}_{L2} + I_{L2}\hat{d}_2\,, \tag{9}$$

where a hat denotes a small-signal perturbation; $D_2 = V_{\text{o,ref}}/V_{\text{ref}}$, and $I_{L2} = P_o/V_{\text{o,ref}}$ are the steady-state duty ratio and load-inductor current, respectively. The load converter employs the same cascaded voltage–current control structure as the baseline source converter, with $C_{i,2} = K_{pc2}$, and $C_{v,2}(s) = K_{pv2} + K_{iv2}/s$. The duty command also includes the output-voltage feedforward term $v_o/V_{\text{ref}}$. Its small-signal duty command is

$$\hat{d}_2 = C_{i,2}\left[C_{v,2}(s)(\hat{v}_{o,ref} - \hat{v}_o) - \hat{i}_{L2}\right] + \frac{\hat{v}_o}{V_{ref}}\,. \tag{10}$$

To evaluate the closed-loop load input impedance $Z_L(s) = \hat{v}/\hat{\imath}$, the load output-voltage reference is held constant, i.e., $\hat{v}_{o,ref} = 0$. Combining (9) and (10) with the linearized $L_2$-$C_2$-$R_2$ load-stage model in Fig. 3 gives

$$Z_L(s) = \frac{\left(sL_2 + V_{ref}C_{i,2}\right)Y_{o,2}(s) + V_{ref}C_{i,2}C_{v,2}(s)}{D_2^{\,2}Y_{o,2}(s) + D_2 I_{L2}\left(V_{ref}^{\;-1} - C_{i,2}\left(C_{v,2}(s) + Y_{o,2}(s)\right)\right)}\,, \tag{11}$$

where $Y_{o,2}(s)=1/R_2+sC_2$. The controller gains are held fixed, whereas $R_2$ is varied to set the operating power $P_o=V_{\text{o,ref}}^2/R_2$. At frequencies well below the load voltage-loop bandwidth, (11) reduces to

$$Z_L(0) = -\frac{V_{ref}^{\;2}}{P_o}\,, \tag{12}$$

demonstrating the negative incremental input resistance at low frequencies. The complete frequency-dependent expression in (11) is used in the subsequent impedance analysis.

TABLE I. Nominal Circuit and Control Parameters

| Source | Value | Load | Value |
|---|---|---|---|
| $V_{\text{in}}$ | 48 V | $V_{\text{o,ref}}$ | 7 V |
| $V_{\text{ref}}$ | 24 V | $P_o$ | 100 W |
| $L$ | 288 μH | $L_2$ | 70 μH |
| $C$ | 68 μF | $C_2$ | 25 μF |
| $R$ | 11.52 Ω | $R_2$ | 0.49 Ω |
| $K_{pc}$ | 0.3 A$^{-1}$ | $K_{pc,2}$ | 0.1833 A$^{-1}$ |
| $\tau_i$ | 20 μs | $K_{pv,2}$ | 0.1579 A/V |
| $K_{pv}$ | 0.213628 A/V | $K_{iv,2}$ | 12.89 kA/(V·s) |
| $K_{iv}$ | 272.708 A/(V·s) | $f_{sw,2}$ | 100 kHz |
| $f_{sw}$ | 100 kHz | $f_{c,i,2}$ | 10 kHz |
| $f_{c,v}$ | 500 Hz | $f_{c,v,2}$ | 1 kHz |

## IV. Simulation Verification

### A. Reference-Response Preservation

Fig. 5 compares the analytical and switched responses to a 0-to-24-V source output bus voltage reference step with the downstream Buck voltage reference set to zero. The baseline and MFC analytical traces coincide, and their switched responses closely follow the corresponding analytical transfer-function (TF) responses. This confirms that the additional MFC path changes the source-port impedance without changing the nominal command response.

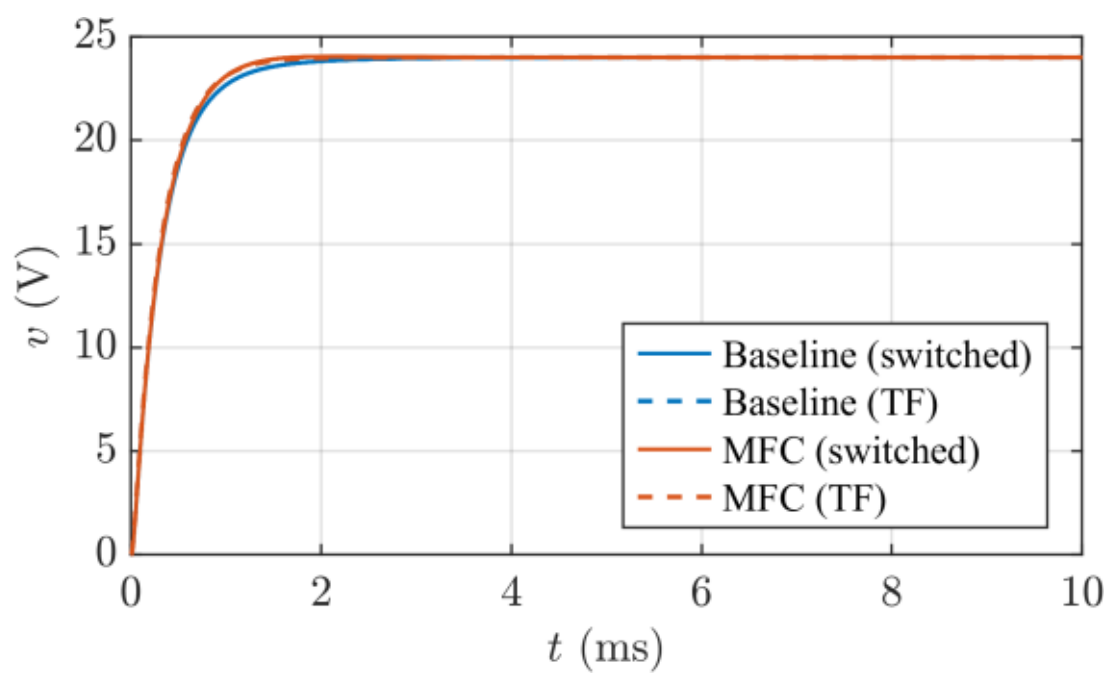

Fig. 5. Source-bus response to a 0-to-24-V reference step with zero downstream voltage reference.

### B. Output-Impedance Reshaping

Fig. 6 compares the baseline and MFC-shaped source output impedances with the closed-loop input impedance of the regulated Buck load at 100 W and at the respective stability limits. Consistent with (5), over the frequency range in which $|1 + T(j\omega)| > 1$, the additional MFC path reduces $|Z_s|$ to $1/|1 + T(j\omega)|$ times its baseline value. As $|T(j\omega)|$ rolls off at higher frequencies, the baseline and MFC-shaped source impedances converge. The resulting impedance reduction shifts the critical source–load impedance intersection from 307.3 Hz at 172.63 W to 655.8 Hz at 254.92 W. Accordingly, the predicted maximum stable load power increases by 82.29 W, corresponding to an improvement of 47.7%.

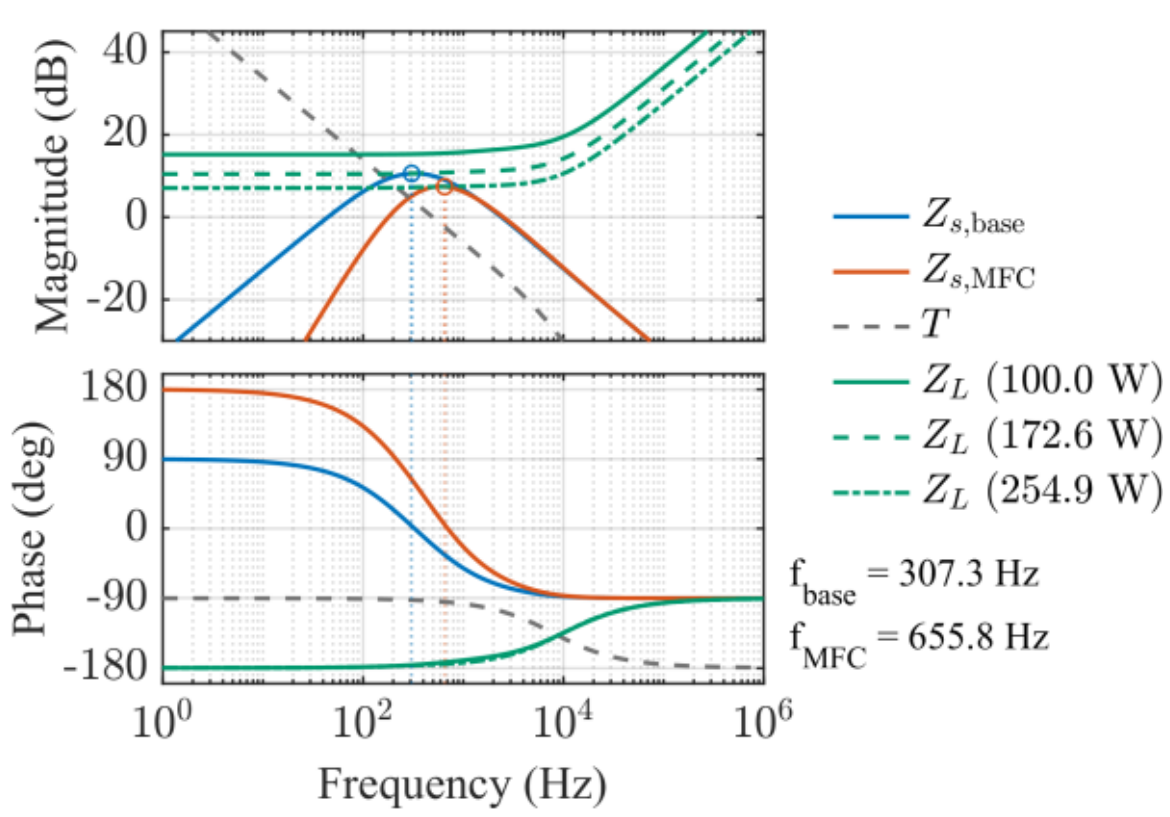

Fig. 6. Source loop gain, source output impedance, and regulated-Buck input impedance at 100 W and the analytical critical powers.

### C. Switched-Model Verification of Stability Improvement

Fig. 7 compares the bus-voltage responses to a load-power step from 175 to 176 W. The baseline-controlled system develops growing oscillations at 176 W, whereas the MFC-controlled system remains stable. As shown in Fig. 8, the MFC-controlled system remains stable at 263 W but develops growing oscillations at 264 W. The resulting stability-boundary brackets are compared with the averaged-model predictions in Table II. The dominant oscillation frequencies at the first unstable operating points are approximately 310 and 667 Hz, respectively, close to the predicted values of 307.3 and 655.8 Hz.

The small differences between the averaged predictions and switched-model brackets may arise from switching effects and nonlinearities omitted by the averaged model, together with the 1-W boundary-search resolution; nevertheless, the two evaluations agree within 3.6% and consistently indicate an approximately 50% increase in the maximum stable load power.

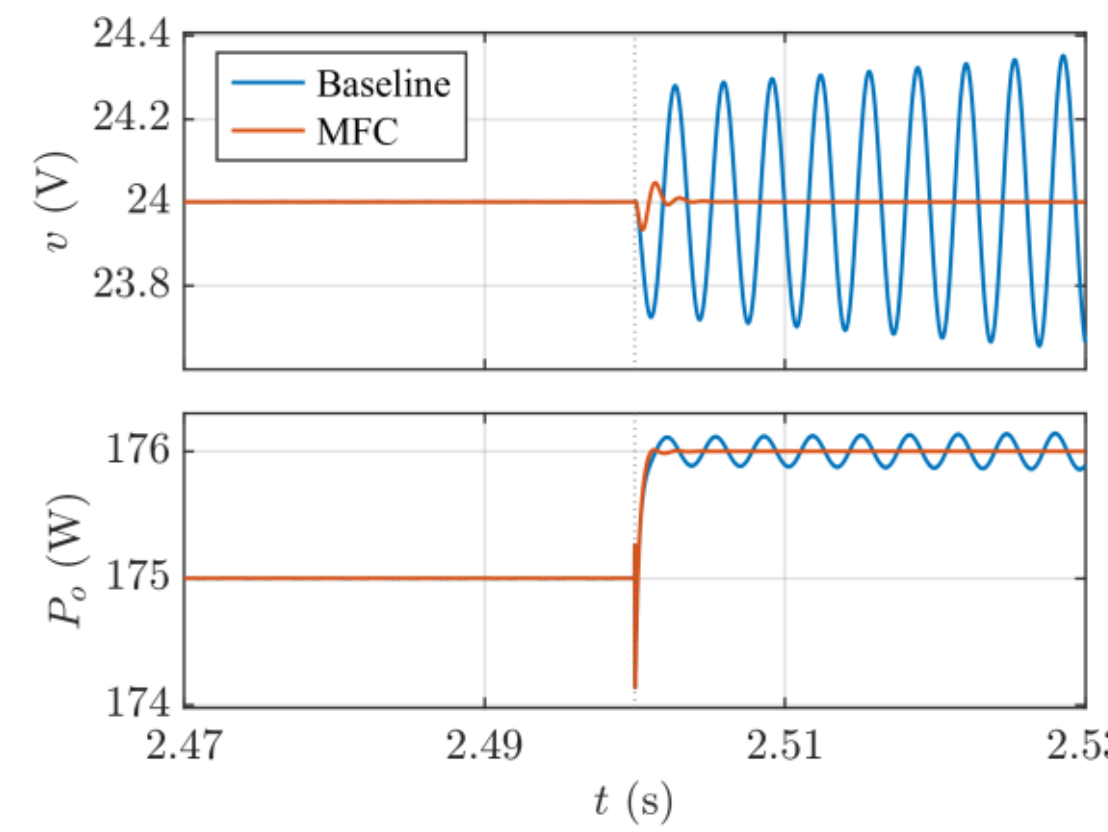

Fig. 7. Switched-model bus-voltage responses to a load-power step from 175 to 176 W under the baseline and MFC controllers.

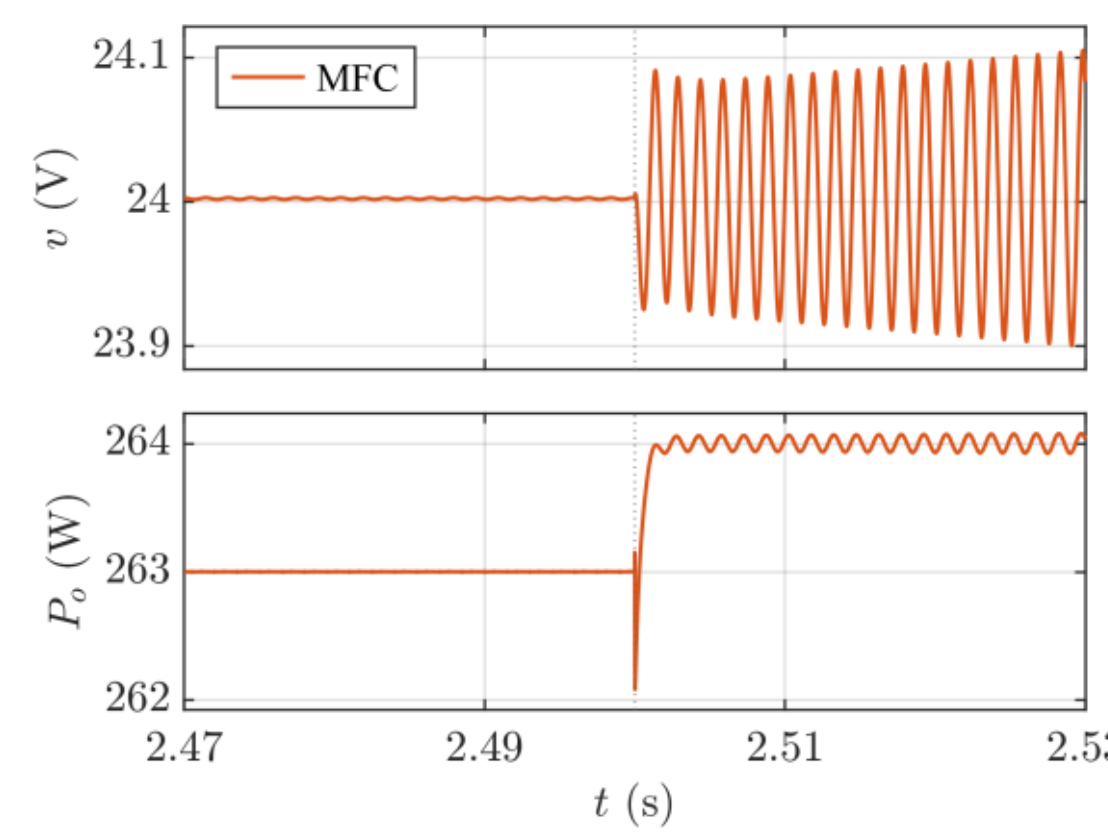

Fig. 8. Switched-model bus-voltage response with MFC for a load-power step from 263 to 264 W.

### D. Large-Signal Regulated-Load Power-Step Responses

Fig. 9 compares the baseline and MFC-controlled systems under the same load-resistance step, which increases the downstream load power from 100 to 176 W. The final power exceeds the predicted baseline stability limit of 172.63 W but remains below the MFC stability limit of 254.92 W. The baseline system exhibits sustained oscillations and fails to settle, whereas the MFC-controlled system restores the dc-bus voltage to 24 V and settles at a downstream load power of 176 W. This comparison shows that the proposed output-impedance reshaping improves the large-signal stability of the cascaded system.

TABLE II. COMPARISON OF LOAD-POWER STABILITY BOUNDARIES

| Evaluation | Baseline Boundary (W) | MFC Boundary (W) | Increase (%) |
|---|---|---|---|
| Averaged-model prediction | 172.63 | 254.92 | 47.7 |
| Switching-model bracket | 175–176 | 263–264 | ≈50 |

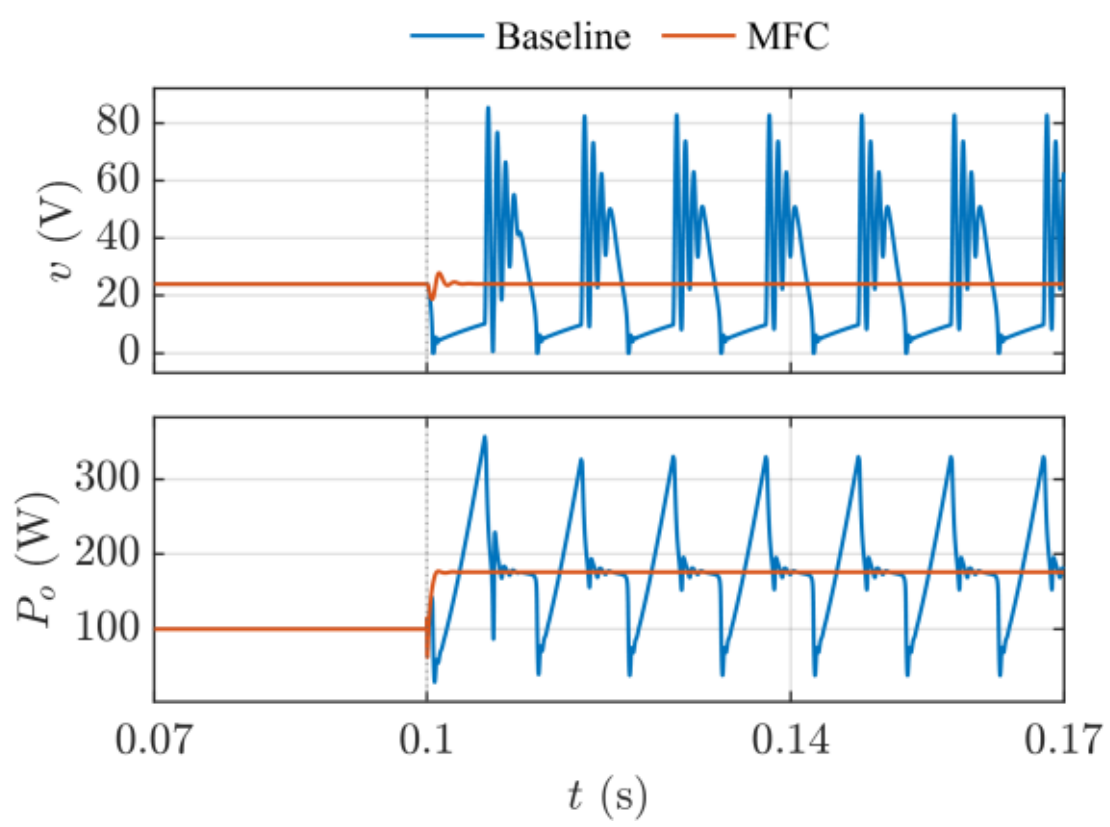

Fig. 9. Switched-model responses to a large-signal load-power step from 100 to 176 W under baseline and MFC control.

## CONCLUSIONS

This paper presents the MFC outer loop as a source-side output-impedance-shaping mechanism for cascaded dc systems. With $C_o(s) = 1+T(s)$, MFC preserves the nominal reference-to-output transfer function and applies $1/[1+T(s)]$ to the source output impedance and the source-load minor loop. Frequency-domain analysis and switched-model simulations of the Buck-CPL case verify the predicted impedance reshaping and associated stability enhancement.

## REFERENCES

[1] W. Du, J. Zhang, Y. Zhang, and Z. Qian, "Stability criterion for cascaded system with constant power load," *IEEE Trans. Power Electron*., vol. 28, no. 4, pp. 1843–1851, Apr. 2013, doi: 10.1109/TPEL.2012.2211619.

[2] A. Emadi, A. Khaligh, C. H. Rivetta, and G. A. Williamson, "Constant power loads and negative impedance instability in automotive systems: Definition, modeling, stability, and control of power electronic converters and motor drives," *IEEE Trans. Veh. Technol*., vol. 55, no. 4, pp. 1112–1125, Jul. 2006.

[3] R. D. Middlebrook, "Input filter considerations in design and application of switching regulators," in *Proc. IEEE Ind. Appl. Soc. Annu. Meeting*, 1976, pp. 366–382.

[4] X. Feng, J. Liu, and F. C. Lee, "Impedance specifications for stable dc distributed power systems," *IEEE Trans. Power Electron*., vol. 17, no. 2, pp. 157–162, Mar. 2002.

[5] X. Liu, A. J. Forsyth, and A. M. Cross, "Negative input-resistance compensator for a constant power load," *IEEE Trans. Ind. Electron*., vol. 54, no. 6, pp. 3188–3196, Dec. 2007.

[6] X. Zhang, X. Ruan, and Q.-C. Zhong, "Improving the stability of cascaded dc/dc converter systems via shaping the input impedance of the load converter with a parallel or series virtual impedance," *IEEE Trans. Ind. Electron*., vol. 62, no. 12, pp. 7499–7512, Dec. 2015.

[7] M. Wu and D. D.-C. Lu, "A novel stabilization method of LC input filter with constant power loads without load performance compromise in dc microgrids," *IEEE Trans. Ind. Electron*., vol. 62, no. 7, pp. 4552–4562, Jul. 2015.

[8] A. M. Rahimi and A. Emadi, "Active damping in dc/dc power electronic converters: A novel method to overcome the problems of constant power loads," *IEEE Trans. Ind. Electron*., vol. 56, no. 5, pp. 1428–1439, May 2009.

[9] Y. Li, P. Jia, and T. Q. Zheng, "Active damping method to reduce the output impedance of the dc–dc converters," *IET Power Electron*., vol. 8, no. 1, pp. 88–95, Jan. 2015.

[10] B. He, W. Chen, X. Li, L. Shu, and X. Ruan, "A power adaptive impedance reshaping strategy for cascaded dc system with buck-type constant power load," *IEEE Trans. Power Electron*., vol. 37, no. 8, pp. 8909–8920, Aug. 2022.

[11] W. Li, Y. Zhang, H. Yuan, X. Zhang, J. Zhu, and S. Li, "A decoupling framework for ultra-local-model-based model-free control of power electronic systems," *Control Eng. Pract*., vol. 177, Art. no. 107196, 2026, doi: 10.1016/j.conengprac.2026.107196.

[12] R. W. Erickson and D. Maksimović, Fundamentals of Power Electronics, 3rd ed. Cham, Switzerland: Springer, 2020.

[13] X. Zhang, J. Xu, J. Wu, B. Bao, G. Zhou, and K. Zhang, "Reduced-order mapping and design-oriented instability for constant on-time current-mode controlled buck converters with a PI compensator," *J. Power Electron*., vol. 17, no. 5, pp. 1298–1307, Sep. 2017.